\documentclass[12pt]{article}

\newcommand{\beq}{\begin{equation}}
\newcommand{\eeq}{\end{equation}}
\newcommand{\beqa}{\begin{eqnarray}}
\newcommand{\eeqa}{\end{eqnarray}}
\newcommand{\nn}{\nonumber}

\def\ifmath#1{\relax\ifmmode#1\else$#1$\fi}
\def\to{\rightarrow}

\def\gsim{{~\raise.15em\hbox{$>$}\kern-.85em
          \lower.35em\hbox{$\sim$}~}}
\def\lsim{{~\raise.15em\hbox{$<$}\kern-.85em
          \lower.35em\hbox{$\sim$}~}}

\def\L{{\cal L}}
\def\s{\!\!\!/}
\def\S{\!\!\!\!/}
\def\L{\Lambda_{\rm QCD}}
\def\P{{\rm P}}
\def\dinal{_{/\!/}}
\def\trans{_\perp}

\begin{document}


\title{\bf\Large Heavy-to-Light Form Factors in the\\
Final Hadron Large Energy Limit of QCD}
\author{J. Charles, A. Le Yaouanc, L. Oliver, O. P\`ene and J.-C. Raynal}
\date{December 14, 1998 (revised January 1999)}
\maketitle

\begin{center}
Laboratoire de Physique Th\'eorique et Hautes \'Energies~\footnote{Laboratoire
associ\'e au Centre National de la Recherche Scientifique - URA D00063\\
E-mail: {\tt Jerome.Charles@th.u-psud.fr}}
\\
Universit\'e de Paris-Sud, B\^atiment 210, F-91405 Orsay Cedex, France
\end{center}

\setcounter{page}{1}

\vskip 5 mm
\begin{flushright}
LPTHE-Orsay 98-77\\ {\tt hep-ph/9812358}
\end{flushright}

\begin{abstract}
We argue that the Large Energy Effective Theory (LEET), originally proposed by
Dugan and Grinstein, is applicable to exclusive semileptonic, radiative and
rare heavy-to-light transitions in the region where the energy release $E$ is
large compared to the strong interaction scale and to the mass of the final
hadron, i.e. for $q^2$ not close to the zero-recoil point. We derive the
Effective Lagrangian from the QCD one, and show that in the limit of heavy
mass $M$ for the initial hadron and large energy $E$ for the final one,
the heavy and light quark fields
behave as two-component spinors. Neglecting QCD short-distance corrections,
this implies that there are only three form factors describing all the
pseudoscalar to pseudoscalar or vector weak current matrix elements. We argue
that the dependence of these form factors with respect to $M$ and $E$ should
be factorizable, the $M$-dependence ($\sqrt{M}$) being derived from the usual
heavy quark expansion while the $E$-dependence is controlled by the behaviour
of the light-cone distribution amplitude near the end-point $u\sim 1$. The
usual expectation of the $\sim (1-u)$ behaviour leads to a $1/E^2$ scaling
law, that is a dipole form in $q^2$. We also show explicitly that in the
appropriate limit, the Light-Cone Sum Rule method satisfies our general
relations as well as the scaling laws in $M$ and $E$ of the form factors, and
obtain very compact and simple expressions for the latter. Finally we note
that this formalism gives theoretical support to the quark model-inspired
methods existing in the literature.
\end{abstract}

\newpage

\section{Introduction}
Nowadays, $|V_{cb}|$ is the third most accurately measured
Cabibbo-Kobayashi-Maskawa (CKM) matrix element, and is quoted by the {\it
Review of Particle Physics}~\cite{PDG} with less than 5\% relative
uncertainty. The Isgur-Wise symmetry~\cite{IW}, and the Heavy Quark Effective
Theory (HQET) description of heavy-to-heavy semileptonic decays, have
permitted such a great success in heavy quark physics. Unfortunately, the HQET
constraints on heavy-to-light decays are quite weak in their original form,
and still do not allow a clean extraction of $|V_{ub}|$ from the present and
future experimental data. The latter CKM coupling has currently a relative
uncertainty of order 25\%, depending on which model is used to evaluate the
hadronic matrix elements~\cite{PDG}. It is thus very important to make
theoretical progress in this field.

The peculiar feature of exclusive heavy-to-light transitions, the prototype of
which is $B\to\pi\ell\nu_{\ell}$, is the large energy $E$ given to the
daughter by the parent hadron, in almost the whole physical phase space except
the vicinity of the zero-recoil point:
\beq
E=\frac{m_B}{2} \left[ 1-\frac{q^2}{m_B^2}+\frac{m_\pi^2}{m_B^2} \right]
\eeq

As we shall see, one may assume that such transitions are dominated by soft
gluon exchange, i.e. the Feynman mechanism, which is not power suppressed with
respect to the hard contribution~\footnote{We neglect in the whole paper
Sudakov effects, which are expected not to induce a large suppression at the
physical $m_B$ scale.}, and does not suffer the $\alpha_s$ suppression
characteristic of the latter~\cite{CZ,BD}; this has to be contrasted to the
case of the pion electromagnetic form factor at very large $Q^2=-q^2$. Then in
the heavy-to-light case, the final active quark should carry most of the
momentum of the light hadron, and the fast degrees of freedom become
essentially classical. The Large Energy Effective Theory (LEET), originally
introduced by Dugan and Grinstein~\cite{DG}, should be the correct tool to
study such transitions: it could provide an Operator Product Expansion the
small parameter of which is $1/E$. As for the initial heavy quark, the
assumption of the soft contribution dominance leads to an expansion in powers
of the inverse heavy mass $M$, based on HQET. Our first result is that {\it to
leading order in $1/M$, $1/E$ and neglecting short-distance QCD corrections,
all the weak current $P\to P(V)$ matrix elements can be expressed in terms of
only three universal form factors.} This implies relations between the usual
semileptonic and penguin form factors which resemble somehow to the well-known
Isgur-Wise relations in heavy-to-heavy transitions.

Then an interesting question is the dependence of these form factors with
respect to the large mass $M$ and the large energy $E$. From the usual heavy
mass expansion of the initial hadron state, we obtain a factorization formula
$\sim\sqrt{M}\,z(E)$. The asymptotic expansion of $z(E)$ is controlled by the
behaviour of the light-cone distribution amplitude of the final hadron near the end-point $u\sim
1$. The usual assumption of the $\sim(1-u)$ behaviour leads to a $z(E)\sim
1/E^2$ scaling law, which implies a $\sim M^{-3/2}/(1-q^2/M^2)^2$ dipole form
for the three universal form factors~\footnote{The $q^2$-dependence of the
form factors in the standard parametrization is discussed in
Section~\ref{pheno}.}.

A strong support in favour of the HQET/LEET formalism for heavy-to-light form
factors is given by the Light-Cone Sum Rules: indeed using the work of several
groups~\cite{CZ}~\cite{khodja1}-\cite{khodja2} we show explicitly that the
latter method automatically satisfies the above relations and scaling laws. In
addition we show in another paper~\cite{BT} that the quark models based on the
Bakamjian-Thomas formalism, which were shown to be covariant and to fulfill
the Isgur-Wise relations in the heavy-to-heavy case~\cite{BT0}, become also
covariant in the $M\to\infty$ and $E\to\infty$ limit. In agreement with our
general results, these models do predict that there are only three independent
form factors in heavy-to-light transitions, and that they scale like
$\sqrt{M}\,z(E)$.

The paper is organized as follows: in Section~\ref{LEET}, we argue on the
validity of LEET for the description of the Feynman mechanism and derive the
correct form of the Lagrangian, as there is a subtlety concerning Dirac
matrices which was missed in the literature. In Section~\ref{zeta} we show
that the use of the HQET and LEET effective quark fields leads to express all
the heavy-to-light ground state form factors in terms of only three universal
functions. The asymptotic $M$- and $E$-dependence of these functions are
discussed, and a $\sim\sqrt{M}/E^2$ form is shown to be the most plausible.
Then in Section~\ref{LCSR} we derive explicitly the $M\to \infty$ and $E\to
\infty$ limit of the Light-Cone Sum Rule formul\ae\ for the weak current
matrix elements and give the universal form factors in terms of $\sqrt{M}/E^2$
times integrals depending on the light-cone distribution amplitudes and the
sum rule parameters, in a very simple and compact way. Finally in
Section~\ref{pheno} we discuss the relation between the matrix elements
parametrized in the standard way and the HQET/LEET universal form factors, and
compare our results with previous approaches that were based on the
constituent quark model.
\section{The LEET Effective Theory}
\label{LEET}
LEET was introduced by Dugan and Grinstein~\cite{DG} to study factorization of
non-leptonic matrix elements in decays like $B\to D^{(*)}\pi,\
D^{(*)}\rho...$, where the light meson is emitted by the $W$-boson. In this
case, both quarks constituting the light energetic meson are fast. However,
Aglietti {\it et al.}~\cite{agliettiNew} have recently argued that such a
situation could not be described by LEET, as the relative transverse momentum
of the fast quarks may be hard. They proposed to use instead the
$\overline{\rm LEET}$ effective theory, a variant of LEET which takes into
account hard transverse degrees of freedom. This seems to be similar to the
description of the heavy quark systems: HQET is the appropriate theory for the
heavy-light hadrons, while NRQCD (Non-Relativistic QCD) should be used for the
quarkonia. Conversely, Aglietti {\it et al.} found that LEET could be used in
semi-inclusive non-leptonic decays such as $B\to DX_u$, where factorization
should hold at the leading order~\cite{agliettiNew}.

Note also that the quark propagator in the LEET limit has gained further
interest with the proposal of the Rome group~\cite{rome} to use it to extract
from the lattice the shape function for semileptonic inclusive $B$-decays, the
structure functions in deep inelastic scattering and the light-cone
distribution amplitudes for exclusive hard processes.

However, no effort seems to have been devoted up to now to investigating how
LEET could be used in processes where only one quark is fast and the other
partons are soft~\footnote{A potential problem of the LEET effective theory is
the so-called ``instability'' phenomenon~\cite{agliettiNew,agliettiOld}:
indeed Aglietti argues in Ref.~\cite{agliettiOld} that the interaction of a
LEET quark with a {\it massless} soft quark ---that constitutes precisely the
case that we are interested in--- generates divergences in the forward
direction. However the mass of a quark in a bound state should rather be
viewed as an {\it off-shellness} of order $\L$, and it can easily be checked
that this instability problem in Aglietti's argument does not occur for a
non-vanishing mass of the soft quark~\cite{martinelli}.}, similarly to the
HQET description of heavy-to-heavy transitions. In this work, we will study
amplitudes where an energetic hadron is connected to the decaying heavy hadron
by a current operator, i.e. exclusive semileptonic, radiative and rare
heavy-to-light decays, such as $B\to\pi\ell\nu_{\ell}$, $B\to K^\ast\gamma$
and $B\to K\ell^+\ell^-$. In the large recoiling region, i.e. for $q^2$
sufficiently far away from the zero-recoil point (of course the physical $q^2$
in radiative decays is exactly zero), the final active quark carries a large
energy (in the rest frame of the parent hadron) and interacts mostly with soft
degrees of freedom ---the spectator quark and the gluons. Thus one may expect
that the trajectory of the fast quark suffers only small fluctuations around
the classical, almost light-like, worldline of the daughter hadron. Actually
there are also hard gluon exchange contributions, through which the large
momentum is shared by both the active and spectator quarks. However the
perturbative calculation of these diagrams typically leads to very small
values compared to the dominant overlap diagram, due to the hard $\alpha_s$
suppression~\cite{CZ,BD}. We will assume that these contributions are
negligible, and we neglect also other radiative corrections for simplicity.

Let us now define the Large Energy Effective Theory in a more systematical
way. From now on, we will refer to high energy exclusive heavy-to-light
decays, and consider only the ground state mesons. The appropriate kinematical
variables for such decays are
\begin{itemize}
\item The four-momentum $p$, mass $M$ and four-velocity $v$ of the heavy hadron
\beq\label{v}
p\equiv Mv
\eeq
\item The four-vector $n$ and the scalar $E$ defined by
\beq
p^\prime\equiv En\,,\ \ \ \ \ v\cdot n\equiv 1
\eeq
where $p^\prime$ is the four-momentum of the light hadron,
$p^{\prime\,2}=m^{\prime\,2}$. Thus
\beq\label{E}
E=v\cdot p^\prime
\eeq
is just the energy of the light hadron in the rest frame of the heavy hadron.
\end{itemize}
In the following we will consider the limit of heavy mass for the initial hadron and
large energy for the final one:
\beq
(\L,m^\prime)\ll (M,E)\,,\ \ \ \ \ \mbox{with $v$ and $n$ fixed.}
\eeq
Note that we do not assume anything for the ratios $E/M$ and $\L/m^\prime$. As
$n^2=m^{\prime\,2}/E^2\to 0$, $n$ becomes light-like in the above limit. In
the rest frame of $v$, with the $z$ direction along $p^\prime$, one has simply
\beq
v=(1,0,0,0)\,,\ \ \ \ \ n\simeq(1,0,0,1)\,.
\eeq
In a general frame one has the normalization conditions
\beq\label{norm}
v^2=1\,,\ \ \ \ \ v\cdot n =1\,,\ \ \ \ \ n^2\simeq 0\,.
\eeq

In a decay like $B\to\pi$, not too close from $q^2=q^2_{\rm
max}=(m_B-m_\pi)^2$, the final active quark gets a very large energy and
should form with the spectator a hadron of finite mass. Thus, neglecting as
said above hard spectator effects, the momentum $r$ of the active quark is
close to the momentum of the hadron:
\beq\label{limite}
r=En+k\,,\ \ \ \ \ \mbox{with $k\sim\L\ll E$ is the residual momentum.}
\eeq

Our goal is now to derive the LEET Lagrangian from the QCD one in the
limit~(\ref{limite}). We would like to separate the large components of the
quark field from the small ones which, corresponding to the negative energy
solutions, should be suppressed by $1/E$. To this aim we follow closely the
simple demonstration given in Ref.~\cite{neubert} for the derivation of the
HQET Lagrangian. We define the projectors
\beq
\P_+=\frac{n\s v\s}{2}\,,\ \ \ \ \ \P_-=\frac{v\s n\s}{2}
\eeq
which indeed verify from Eq.~(\ref{norm})
\beq
\P_\pm^2=\P_\pm\,,\ \ \ \ \ \P_\pm \P_\mp=0\,,\ \ \ \ \ \P_++\P_-=\frac{\{n\s,v\s\}}{2}=1\,.
\eeq

From the full, four-component, quark field $q(x)$ one may define two
two-component projected fields $q_\pm(x)$ by
\beq
q_\pm(x)\equiv e^{iEn\cdot x}\P_\pm q(x)
\eeq
Thus from the projector properties one has
\beq
q(x)=e^{-iEn\cdot x}\left[q_+(x)+q_-(x)\right]
\eeq
with
\beq
\P_\pm q_\pm = q_\pm\,,\ \ \ \ \ \P_\mp q_\pm = 0
\eeq
and
\beq
\overline{q}_\pm\P_\mp=\overline{q}_\pm\,,\ \ \ \ \ \overline{q}_\pm\P_\pm=0\,.
\eeq
It follows immediately that the QCD Lagrangian for the quark $q$, ${\cal
L}_{\rm QCD}=\overline{q}(iD\S-m_q)q$ can be expressed in terms of the $q_\pm$
fields:
\beq
{\cal L}_{\rm QCD}=\overline{q}_+v\s in\cdot Dq_++\overline{q}_+(iD\S-m_q)q_-
+\overline{q}_-(iD\S-m_q)q_++\overline{q}_-v\s(2E+2iv\cdot D-in\cdot D)q_-
\eeq
The equation of motion $(iD\S-m_q)q(x)=0$, projected by $\P_\pm$, reads
\beqa
v\s in\cdot Dq_++\left[iD\S-m_q-v\s (2iv\cdot D-in\cdot D)\right]q_- &=&0\,,\\
(iD\S-m_q-v\s in\cdot D)q_++v\s (2E+2iv\cdot D-in\cdot D)q_-&=&0\,.
\eeqa
The latter equation can formally be solved to express $q_-$ in terms of $q_+$:
\beq
q_-(x)=(2E+2iv\cdot D-in\cdot D+i\epsilon)^{-1}v\s(iD\S-m_q-v\s in\cdot
D)q_+(x)\,.
\eeq
Thus the field $q_-(x)$, corresponding to the negative energy
solutions~\footnote{since $\P_\pm=\frac{1}{2}(1\pm\alpha_z)$ in the rest frame
of $v$. In this frame, the projectors $\P_{\pm}$ coincide with the
$\pm\gamma_3\gamma_0/2$ projectors that are useful for the light-cone
formalism~\cite{BL}. Another possibility is to define
$\P_+=\gamma_\ast\gamma_\bullet/2$ and $\P_-=\gamma_\bullet\gamma_\ast/2$ with
$\gamma_\ast=n\s/n\cdot x$ and $\gamma_\bullet=x\s$: this makes apparent the
resemblance with the light-cone formalism of Ref.~\cite{BF} and leads to the
Lagrangian ${\cal L}_{\rm LEET}=\overline{q}_n[x\s/n\cdot x]in\cdot Dq_n$. The
latter exhibits the invariance under the collinear conformal group as defined
in Ref.~\cite{BF}.} is of order $1/E$ with respect to $q_+(x)$. Physically
this means that the pair creation is suppressed in the effective theory.

To summarize we have obtained the result:
\beq\label{QCD->LEET}
{\cal L}_{\rm QCD}={\cal L}_{\rm LEET}+{\cal O}(1/E)
\eeq
with
\beq\label{L_LEET}
{\cal L}_{\rm LEET}=\overline{q}_nv\s in\cdot Dq_n,
\eeq
where we have defined $q_n(x)\equiv q_+(x)$ to recall the usual notation
$h_v(x)$ for the effective field of HQET. In addition the two-component field
$q_n(x)$ verifies the projection condition
\beq\label{projection}
q_n(x)=\frac{n\s v\s}{2}q_n(x)
\eeq
which implies in particular $n\s q_n=0$. The LEET equation of motion is just
\beq
n\cdot D q_n(x)=0\,.
\eeq

Note that in the literature~\cite{DG,agliettiNew,agliettiOld}, the LEET
Lagrangian was quoted without the $v\s$ factor: indeed these authors have
inferred the Lagrangian from the large energy limit of the QCD quark
propagator. However the limit of the propagator is not sufficient to define
the effective field from the QCD field. We will see that the $v\s$ factor and
the projection condition~(\ref{projection}) have important consequences on the
symmetries of the effective theory and on the constraints on the form factors.

Note also that the assumption of a massless quark, or even a light quark
(compared to $\L$), is not needed to write Eq.~(\ref{QCD->LEET}). The mass
term $\overline{q}_nm_qq_n$ just vanishes because of the
projector~(\ref{projection}). As far as masses are concerned, we only need
$m_q\ll E$ for the quark, and $m^\prime\ll E$ for the hadron, in order for $n$
to become a light-like vector. Of course in phenomenological applications we
will use LEET mainly for the light $u$, $d$ and $s$ quarks. However it is
worth noting that if the $b$ quark were much heavier, say $\sim 20$ GeV, the
heavy $c$ quark in the $B\to D$ transition around $q^2=0$ would have to be
described by LEET rather than by HQET, as the latter theory fails for
$w=E/m^\prime$ too large~\cite{neubert}.

Let us now discuss the global symmetries of LEET: the simplest one is the
flavour symmetry, as there is no mass term in the Lagrangian, meaning that
mass effects should be small for energetic particles.

It is also immediately apparent that the LEET Lagrangian~(\ref{L_LEET}) as
well as the projection condition~(\ref{projection}) are invariant under the
chiral transformation
\beq
q_n(x)\to e^{i\alpha\gamma^5/2}q_n(x)\,.
\eeq
As for massless quarks, it is straightforward to show that the helicity
operator of the LEET quark is just $\gamma^5/2$. This feature, and the fact
that there is no dynamical Dirac matrix in the LEET Lagrangian (coupled to the
covariant derivative $D^\mu$), should indicate that the U(1) chiral symmetry
can be embedded in a larger symmetry group~\cite{su2}. This is actually what
happens. One defines~\footnote{We use $\vec\Sigma\equiv
\left(\begin{array}{cc}\vec\sigma&0\\0&\vec\sigma\end{array}\right)$, where $\vec\sigma$
are the Pauli matrices.} in the rest frame of $v$
\beq
S^1\equiv\frac{1}{2}\gamma^0\Sigma^1=\frac{1}{2}\,\gamma^1\gamma^5\,,\ \ \ \ \
S^2\equiv\frac{1}{2}\,\gamma^0\Sigma^2=\frac{1}{2}\,\gamma^2\gamma^5\,,\ \ \ \
\ S^3\equiv\frac{1}{2}\,\Sigma^3=\frac{1}{2}\,\gamma^5\gamma^0\gamma^3\,.
\eeq
In a general frame, one defines two four-vectors $e^1$ and $e^2$ transverse to
both $v$ and $n$ and
\beq\label{groupe}
S^1=\frac{\gamma^5e\s^1}{2}\,,\ \ \ \ \ S^2=\frac{\gamma^5e\s^2}{2}\,,\ \ \ \
\ S^3=\frac{\gamma^5}{2}(1-v\s n\s)\,.
\eeq
Thus $S^3q_n=\gamma^5q_n$ because of the projector, as said above. As the
$\Sigma^i$ generate the SU(2) group, and from
$\left[\gamma^0,\Sigma^i\right]=0$ and $(\gamma^0)^2=1$, the $S^i$ operators
also verify the SU(2) algebra:
\beq
\left[S^i,S^j\right]=i\epsilon_{ijk}S^k\,.
\eeq
Finally, it is simple to check that both the Lagrangian~(\ref{L_LEET}) and the
projection condition~(\ref{projection}) remain invariant under the
transformation generated by $\vec S$
\beq
q_n(x)\to e^{i\vec\alpha\cdot\vec S}q_n(x)\,.
\eeq
Our conclusion is that to leading order, LEET has as much global symmetry as
HQET, that is a flavour and SU(2) symmetry~\footnote{Note the important r\^ole
played by the $v\s$ factor in the Lagrangian.}.

Unfortunately, unlike the HQET case, the action of the
generators~(\ref{groupe}) {\it on the physical states} is not straightforward
to find. Indeed, the HQET symmetry group holds whatever the internal
kinematical configuration of the heavy hadron~\footnote{except for exchange of
hard momenta ($k\gsim M$), which generates logarithmic corrections that are
calculable in perturbation theory.}, and this is obviously not the case for
LEET. In short, the physical states are not dominated by the ``objects'' that
are expected to be described by LEET (for example, the LEET symmetry does not
imply $m_\rho=m_\pi$). However this problem, that we leave for further
investigation, will not prevent us to find very significant results by
sticking to the quark current operators and simply replacing the QCD light
quark field by the LEET one wherever it is justified, as we will see in
Section~\ref{zeta}.

As for the space-time symmetry, it can be checked that while the HQET
Lagrangian is invariant under the rotation group (more precisely the little
group of $v$), the LEET Lagrangian~(\ref{L_LEET}) is invariant under the group
of the collinear conformal transformations~\footnote{These are defined in
Ref.~\cite{BF}, within the context of the quantization on the light-cone.}.
Furthermore, it should be possible to ``make covariant'' the theory by summing
on the four-vector $n$, similarly to Georgi's procedure~\cite{georgi}
concerning HQET.

Finally the Feynman rules of the LEET effective theory look like the HQET
ones~\cite{neubert}
\beqa
\mbox{LEET quark propagator:} && \frac{iv\s}{n\cdot k+i\epsilon}\,\frac{n\s v\s}{2}\,,\\
\mbox{LEET quark-gluon vertex:} && -ig\,v\s\,T_a\,n^\mu\,.
\eeqa
\section{The Universal Form Factors $\zeta(M,E)$, $\zeta\dinal(M,E)$, $\zeta\trans(M,E)$ and their Scaling Laws}
\label{zeta}
Our purpose in this section is to find the constraints on the heavy-to-light
form factors which may follow from LEET. For definiteness, we consider the
decay of a $B$-meson, although some of our results may apply for $D$-decays.
As for the final particle, $P$ ($V$) stands for a light pseudoscalar (vector)
meson. We are interested in the following matrix elements
\beq\label{MA1}
\left\langle P\left|V^\mu\right|B\right\rangle\,,\ \ \ \ \
\left\langle P\left|T^{\mu\nu}\right|B\right\rangle\,,\nn
\eeq
\beq\label{MA2}
\left\langle V\left|V^\mu\right|B\right\rangle\,,\ \ \ \ \
\left\langle V\left|A^\mu\right|B\right\rangle\,,\ \ \ \ \
\left\langle V\left|T_5^{\mu\nu}\right|B\right\rangle
\eeq
where $V^\mu=\overline{q}\gamma^\mu b$, $A^\mu=\overline{q}\gamma^\mu\gamma_5
b$, $T^{\mu\nu}=\overline{q}\sigma^{\mu\nu}b$ and
$T_5^{\mu\nu}=\overline{q}\sigma^{\mu\nu}\gamma_5 b$ are respectively the
vector, axial, tensor and pseudotensor weak currents, with $q$ the appropriate
active light flavour. The matrix elements
\beq
\left\langle P\left|T_5^{\mu\nu}\right|B\right\rangle\,,\ \ \ \ \
\left\langle P\left|T^{\mu\nu}q_\nu\right|B\right\rangle\,,\nn
\eeq
\beq
\left\langle V\left|T^{\mu\nu}\right|B\right\rangle\,,\ \ \ \ \
\left\langle V\left|T_{(5)}^{\mu\nu}q_\nu\right|B\right\rangle
\eeq
($q_\mu=p_\mu-p^\prime_\mu=Mv_\mu-En_\mu$ is the four-momentum transfer) can
obviously be obtained from Eqs.~(\ref{MA1})-(\ref{MA2}) by using
$\gamma_5\sigma^{\mu\nu}=\frac{i}{2}\epsilon^{\mu\nu\rho\sigma}\sigma_{\rho\sigma}$
and/or contracting with $q_\nu$. Let us recall the relation between $q^2$ and
$E=v\cdot p^\prime$
\beq\label{q2}
q^2=M^2-2ME+m^{\prime\,2}\ \ \Longleftrightarrow\ \
E=\frac{M}{2}\left(1-\frac{q^2}{M^2}+\frac{m^{\prime\,2}}{M^2}\right)\,.
\eeq

As a starting point, we decompose in all generality the above matrix elements
in terms of Lorentz invariant form factors. We adopt a parametrization that is
convenient to study the $M\to\infty$ and $E\to\infty$ limit, i.e. we use the
variables $(v^\mu,n^\mu,E)$ rather than $(p^\mu,p^{\prime\mu},q^2)$. Some
caution is needed to treat the polarization vector $\epsilon^\mu$ of the
vector meson: indeed when $E\to\infty$, one has $\epsilon^\mu\trans={\cal
O}(1)$ for a transverse meson while $\epsilon^\mu\dinal={\cal O}(E/m_V)$ for a
longitudinal one. Thus we decompose the matrix elements on the Lorentz
structures $v^\mu$, $n^\mu$, $\epsilon^{\ast\mu}-(\epsilon^\ast\cdot v)n^\mu$,
$\frac{m_V}{E}(\epsilon^\ast\cdot v)n^\mu$, $\frac{m_V}{E}(\epsilon^\ast\cdot
v)v^\mu$ and $\epsilon^{\mu\nu\rho\sigma}v_\nu n_\rho\epsilon^\ast_\sigma$
which are finite in the asymptotic limit $M\to \infty$ and $E\to \infty$:
\beqa
\left\langle P\left|V^\mu\right|B\right\rangle&=&
2E\left[\zeta^{(v)}(M,E)n^\mu+\zeta^{(v)}_1(M,E)v^\mu\right]\,,\label{matrixDeb}\\
\left\langle P\left|T^{\mu\nu}\right|B\right\rangle&=&
i2E\,\zeta^{(t)}(M,E)\left(n^\mu v^\nu-n^\nu v^\mu\right)\,,\\
\left\langle V\left|V^\mu\right|B\right\rangle&=&
i2E\,\zeta\trans^{(v)}(M,E)\,\epsilon^{\mu\nu\rho\sigma}v_\nu
n_\rho\epsilon^\ast_\sigma\,,\label{B->Vv}\\
\left\langle V\left|A^\mu\right|B\right\rangle&=&
2E\,\zeta\trans^{(a)}(M,E)\left[\epsilon^{\ast\mu}-(\epsilon^\ast\cdot
v)n^\mu\right]\nn\\ &&+2E\frac{m_V}{E}(\epsilon^\ast\cdot v)
\left[\zeta\dinal^{(a)}(M,E)n^\mu+\zeta^{(a)}_{1/\!/}(M,E)v^\mu\right]\,,\label{B->Va}\\
\left\langle V\left|T_5^{\mu\nu}\right|B\right\rangle&=&
-i2E\zeta^{(t_5)}_{1\perp}(M,E)\left\{\left[\epsilon^{\ast\mu}-(\epsilon^\ast\cdot
v)n^\mu\right]v^\nu-
\left[\epsilon^{\ast\nu}-(\epsilon^\ast\cdot v)n^\nu\right]v^\mu\right\}\ \ \ \\
&&-i2E\zeta^{(t_5)}\trans(M,E)\left(\epsilon^{\ast\mu}n^\nu-
\epsilon^{\ast\nu}n^\mu\right)\\
&&-i2E\zeta^{(t_5)}\dinal(M,E)\frac{m_V}{E}(\epsilon^\ast\cdot v)\left(n^\mu
v^\nu-n^\nu v^\mu\right)\,.\label{matrixFin}
\eeqa
Some additional comments on Eqs.~(\ref{matrixDeb})-(\ref{matrixFin}) are in
order~\footnote{We use the usual relativistic normalization of states and
$\epsilon^{0123}=+1$. Note that our phase convention for the vector mesons
differs by a $i$ factor from the one used in
Refs.~\cite{ali,ball2,aliev,ball4,ball3,balltwist4}:
$\left.\left|V\right.\right\rangle_{\rm
This~work}=+i\left.\left|V\right.\right\rangle_{\rm
Ref.~\cite{ali}}$.\label{conventions}}: the $2E$ overall normalization factor
has been chosen to restore the dimensionality and for further convenience (cf.
Eq.~(\ref{f+})). The superscripts $(v)$, $(a)$... refer to the Dirac structure
of the current operators. Furthermore it is clear that for a matrix element to
a longitudinal (respectively transverse) vector meson, only the form factors
with a $/\!/$ (respectively $\perp$) subscript contribute in the $M\to \infty$
and $E\to \infty$ limit. Finally we have made explicit the dependence of the
form factors with respect to $M$ and $E$, although they also depend on
$m^\prime=m_P$ or $m_V$.

Let us now expose our argument, in the $M\to\infty$ and $E\to\infty$ limit. On
the one hand we use LEET to describe the final active quark. Thus the quark
field $\overline{q}(0)$ in the current operators will be replaced by the
effective LEET field $\overline{q}_n(0)$, with $\overline{q}_nv\s
n\s/2=\overline{q}_n$. To exploit the latter equation, two useful Dirac
identities are
\beqa
\frac{v\s n\s}{2}\gamma^\mu&=&\frac{v\s n\s}{2}\left[n^\mu v\s+i\epsilon^{\mu\nu\rho\sigma}v_\nu n_\rho\gamma_\sigma\gamma_5\right]\,,\label{dirac1}\\
\frac{v\s n\s}{2}\sigma^{\mu\nu}&=&\frac{v\s n\s}{2}\left[i\left(n^\mu v^\nu-n^\nu v^\mu\right)
-i\left(n^\mu\gamma^\nu-n^\nu\gamma^\mu\right)v\s
+\epsilon^{\mu\nu\rho\sigma}v_\rho n_\sigma\gamma_5\right]\label{dirac2}\,.
\eeqa

On the other hand, one may wonder about the initial heavy quark: should it be
described by QCD or HQET ? We have already noticed that to leading order LEET
neglects hard spectator effects, and that the hard momenta are ``integrated
out'', leaving only soft, non-perturbative degrees of freedom. Thus for
consistency one should use HQET to describe the initial heavy quark, and
replace the quark field $b(0)$ by $b_v(0)$ with $v\s b_v=b_v$. This supports
the proposed conjecture that HQET may be applied to the whole physical
kinematical range in heavy-to-light semileptonic decays~\cite{BD}, and is not
restricted to the small recoil region, that is the Isgur-Wise limit~\cite{IW}.
Note that this HQET/LEET formalism is not more than a {\it soft contribution
dominance assumption}, that will have short-distance $\alpha_s$ corrections,
and non-perturbative $1/M$ and $1/E$ corrections.

Now we are in position to reduce the number of independent form factors, by
simply replacing the quark current operator $\overline{q}\,\Gamma b$ by the
effective one $\overline{q}_n\Gamma b_v$, which is finite in the $M\to\infty$
and $E\to\infty$ limit. From the constraints $v\s b_v=b_v$ and $n\s q_n=0$ and
from Eqs.~(\ref{dirac1})-(\ref{dirac2}) we find the following relations
between the currents, to leading order in $1/M$ and $1/E$:
\beqa
\overline{q}_n b_v&=&v_\mu \overline{q}_n\gamma^\mu b_v\,,\label{courantsDeb}\\
\overline{q}_n\gamma^\mu b_v&=&n^\mu \overline{q}_nb_v+i\epsilon^{\mu\nu\rho\sigma}v_\nu n_\rho\overline{q}_n\gamma_\sigma\gamma_5b_v\,,\\
\overline{q}_n\gamma^\mu\gamma_5 b_v&=&-n^\mu \overline{q}_n\gamma_5b_v+i\epsilon^{\mu\nu\rho\sigma}v_\nu n_\rho\overline{q}_n\gamma_\sigma b_v\,,\\
\overline{q}_n\sigma^{\mu\nu}b_v&=&i\left[n^\mu v^\nu\overline{q}_n b_v
-n^\mu \overline{q}_n\gamma^\nu b_v-(\mu \leftrightarrow \nu)\right]
+\epsilon^{\mu\nu\rho\sigma}v_\rho n_\sigma\overline{q}_n\gamma_5 b_v\,,\\
\overline{q}_n\sigma^{\mu\nu}\gamma_5 b_v&=&i\left[n^\mu v^\nu\overline{q}_n \gamma_5b_v
+n^\mu \overline{q}_n\gamma^\nu\gamma_5 b_v-(\mu \leftrightarrow \nu)\right]
+\epsilon^{\mu\nu\rho\sigma}v_\rho n_\sigma\overline{q}_n
b_v\,.\label{courantsFin}
\eeqa
Reporting Eqs.~(\ref{courantsDeb})-(\ref{courantsFin}) in
Eqs.~(\ref{matrixDeb})-(\ref{matrixFin}) we find
\beqa
\zeta^{(v)}_1=\zeta^{(a)}_{1/\!/}=\zeta^{(t_5)}_{1\perp}&=&0\,,\\
\zeta^{(v)}=\zeta^{(t)}&\equiv&\zeta\,,\\
\zeta\trans^{(v)}=\zeta\trans^{(a)}=\zeta\trans^{(t_5)}&\equiv&\zeta\trans\,,\\
\zeta\dinal^{(a)}=\zeta\dinal^{(t_5)}&\equiv&\zeta\dinal\,.
\eeqa

Thus {\it to leading order in $1/M$, $1/E$ and $\alpha_s$ there are only three
independent form factors in heavy-to-light $B\to P(V)$ transitions,} which
from now on we will denote by $\zeta$ (for $B\to P$), $\zeta\dinal$ and
$\zeta\trans$ (for $B\to V$). This implies non-trivial relations between the
usual form factors $f_+$, $A_1$ etc. (see Section~\ref{pheno}). Note that
among these relations there are the well-known Isgur-Wise relations~\cite{IW}
between the penguin and semileptonic form factors which follow from $v\s
h_v=h_v$ only, while the relations among the semileptonic form factors
stemming from $v\s h_v=h_v$ {\it and} $\overline{q}_n=\overline{q}_n v\s
n\s/2$ are new (they resemble to Stech's~\cite{stech} and
Soares'~\cite{soares} quark model relations, as we discuss in
Section~\ref{pheno}).

The $\zeta(M,E)$ functions have a simple dependence with respect to the large
mass $M$. Indeed it is well known that the following relation between the QCD
and HQET eigenstates holds in the heavy mass limit~\cite{neubert}:
\beq\label{etatHQET}
\left|\left. B,\,p_\mu \right\rangle\right._{\rm QCD} = \sqrt{M} \left|\left. B,\,v_\mu \right.\right\rangle_{\rm HQET}
\eeq
where $\left|\left. B,\,v_\mu \right\rangle\right._{\rm HQET}$ is independent
of $M$. Thus the matrix elements~(\ref{MA1})-(\ref{MA2}) become
\beq\label{reduc}
\left\langle P\,(V)\left|\,\overline{q}\,\Gamma b\,\right|B,\,p_\mu\right\rangle_{\rm QCD}=
\sqrt{M}\,\left\langle P\,(V)\left|\,\overline{q}_n\Gamma b_v\,\right|B,\,v_\mu\right\rangle_{\rm HQET}
\eeq
where the only Lorentz scalar arising in the covariant decomposition of the
right-hand side matrix element is $E=v\cdot p^\prime$ (and $\epsilon^\ast\cdot
v$ in the $B\to V$ case). From Eqs.~(\ref{reduc})
and~(\ref{matrixDeb})-(\ref{matrixFin}) one has
\beq\label{developpe}
\zeta(M,E)=\sqrt{M}\left[z^{(0)}(E)+\sum_{k=1}^\infty\frac{z^{(k)}(E)}{M^k}\right]\,,
\eeq
where the $z^{(k)}(E)/M^k$, $k\ge 1$, stand for higher order terms stemming
from the HQET Lagrangian and from the heavy-to-light current operator. Similar
$1/M$ expansion apply for $\zeta\dinal$ and $\zeta\trans$.

In the original Isgur-Wise derivation~\cite{IW}, $M$ is sent to infinity while
$E$ is kept fixed, and the well-known $\sim\sqrt{M}$ scaling is obtained for
$E\ll M$. Actually whatever the ratio $E/M$, we may also take $E\to\infty$
{\it provided that the $z^{(k)}(E)$, $k\ge 1$, are not enhanced by powers of
$E$ with respect to $z^{(0)}(E)$}, which is unlikely. Indeed the $z^{(k)}(E)$
are suppressed by some power of the large scale $E$, which is related to the
suppression of the wave function of the light energetic meson when one quark
carries most of the momentum of the hadron, i.e. for the Feynman variable
$u\sim 1$. This suppression is {\it universal}, belongs to the properties of
the final state, and thus should hold for {\it all the operators contributing
to the expansion~(\ref{developpe})}. In the end one obtains a factorized
scaling law for {\it any ratio} $E/M$
\beq\label{scaling}
\zeta(M,E)=\sqrt{M}\,z(E)\,,\ \ \
\zeta\dinal(M,E)=\sqrt{M}\,z\dinal(E)\,,\ \ \
\zeta\trans(M,E)=\sqrt{M}\,z\trans(E)\,.
\eeq
We expect that the potential $\sim E/M$ non-factorizable corrections to
Eq.~(\ref{scaling}) will be suppressed by an additional power of the large
scales $M$ or $E$, or by $\alpha_s$.

The question of the definite asymptotic $E$-dependence is more involved,
because we have no relation for LEET comparable to Eq.~(\ref{etatHQET}).
However it is now well accepted that at $q^2=0$, the Feynman mechanism
contribution to the form factors should behave as the hard one, that is $\sim
M^{-3/2}$~\cite{CZ,ali,ball2,khodja2}, although there is no really rigorous
proof of that. As $E\sim M$ at $q^2=0$, it implies that the $z(E)$ functions
in Eq.~(\ref{scaling}) behave like $\sim 1/E^2$. As already said, this follows
from the behaviour of the final state light-cone wave function near the end-point $u\sim
1$~\cite{CZ}: it is argued in Ref.~\cite{CZphysRep} that the light-cone
twist-two distribution amplitude, renormalized at a low scale $\mu\gsim 1$
GeV, vanishes linearly at $u\sim 1$ like each term of its expansion in the
Gegenbauer polynomials. Integrating over a region shrunk to $\Delta u\sim
1/E$, which is the signature of the Feynman mechanism, one obtains the $1/E^2$
scaling law:
\beq
z(E)\sim\int_{1-\frac{\Lambda}{E}}^{1}du\,(1-u)\sim \frac{1}{E^2}\,.
\eeq
Should the behaviour of the distribution amplitudes near $u\sim 1$ be
different from the usual $\sim(1-u)$ expectation, the power law in $1/E$
should be changed. We stress again that the $\sim 1/E^2$ suppression (or
whatever $\sim 1/E^n$) should affect {\it all} the $z^{(k)}$ functions in
Eq.~(\ref{developpe}). We will see in the next section that this is exactly
what the Light-Cone Sum Rule method predicts.

Here we would like to make an important comment: it is sometimes said that
because the asymptotic $M$-dependence is different at the two particular
points $q^2=0$ and $q^2=q^2_{\rm max}=(M-m^\prime)^2$, HQET could not be valid
in the whole range of $q^2$. However the extended HQET scaling
prediction~(\ref{scaling}) is fully compatible with both the Isgur-Wise
$\sim\sqrt{M}$ scaling for $E\ll M$ and the Chernyak-Zhitnitsky $\sim
M^{-3/2}$ scaling at $q^2=0$ provided $z(E)\sim 1/E^2$ at large $E$. Note also
that the dipole scaling of $z(E)$ at large $E$ does not prevent it to be
pole-like at finite $E$, according to the idea of $B^\ast$ meson dominance.

For clarity, we summarize here our results: in the $M\to \infty$ and $E\to
\infty$ limit, the heavy-to-light weak current matrix elements depend on only
three independent dimensionless form factors $\zeta$, $\zeta\dinal$ and
$\zeta\trans$
\beqa
\left\langle P\left|V^\mu\right|B\right\rangle&=&
2E\,\zeta(M,E)n^\mu\,,\label{res1Deb}\\
\left\langle P\left|T^{\mu\nu}\right|B\right\rangle&=&
i2E\,\zeta(M,E)\left(n^\mu v^\nu-n^\nu v^\mu\right)\,,\\
\left\langle V\left|V^\mu\right|B\right\rangle&=&
i2E\,\zeta\trans(M,E)\,\epsilon^{\mu\nu\rho\sigma}v_\nu
n_\rho\epsilon^\ast_\sigma\,,\\
\left\langle V\left|A^\mu\right|B\right\rangle&=&
2E\left\{\zeta\trans(M,E)\left[\epsilon^{\ast\mu}-(\epsilon^\ast\cdot
v)n^\mu\right] +\,\zeta\dinal(M,E)\frac{m_V}{E}(\epsilon^\ast\cdot
v)n^\mu\right\}\,,\\
\left\langle V\left|T_5^{\mu\nu}\right|B\right\rangle&=&
-i2E\zeta\trans(M,E)\left(\epsilon^{\ast\mu}n^\nu-
\epsilon^{\ast\nu}n^\mu\right)\\
&&-i2E\zeta\dinal(M,E)\frac{m_V}{E}(\epsilon^\ast\cdot v)\left(n^\mu
v^\nu-n^\nu v^\mu\right)\,.\label{res1Fin}
\eeqa
These three universal form factors have a $\sqrt{M}$ dependence
\beq\label{res2}
\zeta(M,E)=\sqrt{M}\,z(E)\,,\ \ \zeta\dinal(M,E)=\sqrt{M}\,z\dinal(E)\,,\ \
\zeta\trans(M,E)=\sqrt{M}\,z\trans(E)\,,
\eeq
and we have argued that a $1/E^2$ dependence is very plausible, thus
\beqa
\zeta(M,E)&=&C\frac{\sqrt{M}}{E^2}=4C\frac{M^{-3/2}}{(1-q^2/M^2)^2}\,,\label{res3Deb}\\
\zeta\dinal(M,E)&=&C\dinal\frac{\sqrt{M}}{E^2}=4C\dinal\frac{M^{-3/2}}{(1-q^2/M^2)^2}\,,\\
\zeta\trans(M,E)&=&C\trans\frac{\sqrt{M}}{E^2}=4C\trans\frac{M^{-3/2}}{(1-q^2/M^2)^2}\,.\label{res3Fin}
\eeqa
where $C$, $C\dinal$ and $C\trans$ are unknown dimensionful constants, of
order $\L^{3/2}$.

Let us repeat however that Eqs.~(\ref{res3Deb})-(\ref{res3Fin}) are not on as
solid grounds as Eqs.~(\ref{res1Deb})-(\ref{res2}). Accepting them
nevertheless, everything but three normalization constants is known about the
form factors, which may constitute an even more favourable situation compared
to the heavy-to-heavy case, at least from the mathematical point of view.
Ironically, a model-independent value of the form factors at some particular
$q^2$ would have been more interesting than the $q^2$-dependence, from the
point of view of the extraction of $|V_{ub}|$.

Before closing this section, we would like to make clear the expected region
of validity of the final hadron large energy limit: from Eq.~(\ref{q2}), it
is not restricted, at least formally, to the small values of $q^2$. Indeed,
$q^2$ can even be ${\cal O}(M^2)$ (e.g. $q^2=\alpha M^2$), provided that
$1-q^2/M^2={\cal O}(1)$ (that is $\alpha\ne 1$) and that $M$ is large enough
to get from Eq.~(\ref{q2}) the conditions $\L\ll E$, $m^\prime\ll E$. However
the region near the zero-recoil point, $q^2=M^2-M\chi$ with $\chi$ finite, is
outside the LEET domain. Of course, the realistic world is much more
complicated, and one may expect sizeable non-LEET effects for physical quark
and hadron masses, as one is going from $q^2=0$ to $q^2=q^2_{\rm max}$. For
$B$-decays, one may hope that LEET will be valid for $0\le q^2\lsim$10--15
GeV$^2$ although a more precise answer cannot be given without a careful study
of the subleading terms.

In addition, we are aware of the fact that Eqs.~(\ref{res1Deb}-\ref{res3Fin})
should receive logarithmic radiative corrections ($\sim \ln(M)$ and
$\sim\ln(E)$)~\cite{DG,gregory}, which might be computed by matching QCD onto
HQET/LEET. These calculations are beyond the scope of this paper.
\section{Light-Cone Sum Rules in the Final Hadron Lar\-ge Energy Limit}
\label{LCSR}
In this Section, we shall show that the  Light-Cone Sum Rule (LCSR) method is
fully compatible with the HQET/LEET formalism that we have discussed above;
moreover we obtain below explicit expressions for the three universal form
factors, that are strikingly simple.

Chernyak and Zhitnitsky~\cite{CZ} were the first authors to use the LCSR
method to calculate the heavy-to-light form factors in the region where the
energy release is sufficiently large (actually at $q^2=0$). The basic idea is
to describe the decaying heavy hadron by an interpolating local current and to
use the quark-hadron duality and the Borel transformation to suppress the
contribution of the excited $B$-states and of the continuum. On the other
hand, the light hadron enters the game through the light-cone distribution
amplitudes, order by order in the twist expansion. Later, the method has been
developed for $q^2\ne 0$ by several groups who have taken into account
higher-twist and radiative corrections effects~\cite{khodja1}-\cite{khodja2}.
It has to be considered as a QCD-based approximation, although, to our
knowledge, there is not a well defined limit of the underlying theory in which
the sum rules are exact.

Our purpose is thus to study the $M\to\infty$ and $E\to\infty$ limit of the
LCSR expressions for the form factors, using mainly the explicit formul\ae\ of
Refs.~\cite{khodja1,aliev,khodja2}. For simplicity we do not consider here the
tensor form factors, only the vector and axial ones. Thanks to the
considerable effort developed in the
literature~\cite{CZ}~\cite{khodja1}-\cite{khodja2}, our calculation does not
pose any major problem and will not be reported here in detail; nevertheless
some comments are in order:
\begin{itemize}
\item
{\it The twist expansion does not match the $1/M$ and $1/E$ power expansion.}
We have checked that the leading order contributions to the $M\to \infty$ and
$E\to \infty$ limit depend not only on the two-particle twist-two but also on
the twist-three distribution amplitudes while higher-twist ($\ge 4$) and
multi-particle ($\ge 3$) distribution amplitudes are power suppressed. This
was already found in Refs.~\cite{CZ,ali,ball2,khodja2} where the heavy quark
expansion was considered at the particular point $q^2=0$. This is also
compatible with the finding of Ref.~\cite{ball1} that twist-three terms are
numerically as important as the twist-two ones, while the twist-four and
multi-particle contributions are much smaller. Actually, for the twist
expansion to make sense, the number of different twists contributing to a
given order in $1/M$ and $1/E$ should be finite, which seems indeed to be the
case.
\item
According to the preceding point, the early calculations of the $B\to V$ form
factors are not consistent from the point of view of the $M\to \infty$ and
$E\to \infty$ limit. Indeed the authors of Refs.~\cite{ali,ball2,aliev} have
not taken into account the contribution of the two-particle twist-three
distribution amplitudes $h^{(t)}\dinal(u)$ and $h^{(s)}\dinal(u)$, as defined
in Ref.~\cite{ball3}, which however contribute to leading order on the same
footing as the ones of twist-two. To our knowledge, only in the recent
works~\cite{ball4} these functions have been considered. We calculate the
corresponding terms below.
\item
{\it The ``surface terms'' should be kept systematically.} These terms come
from the integration by parts after the Borel transformation, as discussed in
the Appendix A of Ref.~\cite{ball2}. However for our purpose we have found
simpler and equivalent to perform the integration by parts {\it before} the
Borel transformation; thus in the calculations below we will use
\beq\label{bypart}
\int_0^1\frac{du}{\Delta^2}\,f(u)=\frac{1}{2}\int_0^1
\frac{du}{\Delta}\,\frac{1}{q\cdot p+um^{\prime\,2}}
\left[\frac{m^{\prime\,2}}{q\cdot p+um^{\prime\,2}}f(u)-f^\prime(u)\right]
\eeq
where $f(u)$ is a function of $u$ which verifies $f(0)=f(1)=0$ and
\beq
\Delta=m_b^2-(q+up^\prime)^2\,.
\eeq
Then the Borel transformation is done with respect to the variable
$(q+p^{\prime})^2$, according to
\beq
\frac{1}{\Delta}\ \to\ \frac{1}{u{\cal M}^2}\,\exp\left[-\frac{m_b^2-(1-u)q^2+u(1-u)m^{\prime\,2}}{u{\cal M}^2}\right]
\eeq
where ${\cal M}^2$ is the Borel parameter.
\item
In order to find the $m_b\to\infty$ limit of the LCSR, the sum rule parameters
have to be rescaled. Following Refs.~\cite{CZ,ali,ball2,khodja2} we write the
Borel parameter
\beq
{\cal M}^2=m_b\mu_0
\eeq
and the continuum threshold
\beq
s^B_0=(m_b+\omega_0)^2\,,
\eeq
where the rescaled parameters $\mu_0$ and $\omega_0$ are finite in the
$m_b\to\infty$ limit.

Moreover we note~\cite{khodja1,aliev,khodja2} that the integration domain
after the Borel transformation and the continuum subtraction is $u\in[u_{\rm
min},\,1]$, where
\beq
u_{\rm min}=\frac{m_b^2-q^2}{s_0^B-q^2}\ \ \ \ \mbox{(for $m^\prime\simeq
0$).}
\eeq
In the $m_b\to\infty$ and $E\to\infty$ limit, one has the expansion $u_{\rm
min}\simeq 1-\omega_0/E$, which shows that this is indeed the size of $E$ that
selects the $u\sim 1$ region.
\end{itemize}

We have now all the elements to perform the calculation. We use the standard
notation for the decay constants $f_B$, $f_P$, $f_V$ anf $f^\perp_V$ and for
the distribution amplitudes~\footnote{All these functions are defined and
discussed at length in Ref.~\cite{ball3}.}, $\phi$, $\phi_p$, $\phi_\sigma$,
$\phi\dinal$, $h\dinal^{(s)}$, $h\dinal^{(t)}$, $\phi\trans$, $g\trans^{(v)}$
and $g\trans^{(a)}$. The behaviour of these functions near $u\to 1$ is assumed
to be {\it identical to each term of their conformal expansion in the
Gegenbauer polynomials}~\cite{CZphysRep}, that is, up to $\sim (1-u)\ln (1-u)$
terms proportional to the light quark masses~\cite{ball3},
\beqa
\phi\sim\phi_\sigma\sim\phi\dinal\sim\phi\trans\sim h\dinal^{(s)}\sim g\trans^{(a)}&\sim& (1-u)\,,\\
\phi_p\sim h\dinal^{(t)}\sim g\trans^{(v)}&\sim& {\rm C^{st}}\,.
\eeqa

For the $B\to P$ transitions we simply use the correlators given by
formul\ae~(17) in Ref.~\cite{khodja1} and~(14) in Ref.~\cite{khodja2} and
perform the Borel transformation after using Eq.~(\ref{bypart}). The standard
procedure to subtract the continuum is applied, and we find that in the limit
$m_b\to\infty$ and $E\to\infty$ the $B\to P$ semileptonic matrix element can
be written under the form~(\ref{matrixDeb}) with
\beqa
\zeta^{(v)}(M,E)&=&\frac{1}{f_B}\frac{1}{2E^2}
\left[-f_P\phi^\prime(1)I_2(\omega_0,\mu_0)-\frac{f_Pm_P^2}{6(m_{q_1}+m_{q_2})}\phi_\sigma^\prime(1)I_1(\omega_0,\mu_0)\right]\,,\label{PSR1}\\
\zeta_1^{(v)}(M,E)&=&\frac{1}{f_B}\frac{1}{2E^2}\frac{f_Pm_P^2}{m_{q_1}+m_{q_2}}
\left[\phi_p(1)+\frac{1}{6}\phi_\sigma^\prime(1)\right]I_1(\omega_0,\mu_0)\,.\label{PSR2}
\eeqa
Here $m_{q_{1,2}}$ stand for the masses of the quarks making the light
pseudoscalar  $q_1q_2$ meson, and the $I_j(\omega_0,\mu_0)$ are functions of
the sum rule parameters $(\omega_0,\mu_0)$ through
\beq
I_j(\omega_0,\mu_0)=\int_0^{\omega_0}d\omega\,\omega^j\exp\left[\frac{2}{\mu_0}\left(
\overline{\Lambda}-\omega\right)\right]\ \ \ \ \ j=1,\,2
\eeq
where $\overline{\Lambda}$ is the binding energy of the heavy meson,
$\overline{\Lambda}=m_B-m_b$, which is finite in the $m_b\to\infty$ limit. For
the particular point $q^2=0$, Eqs.~(\ref{PSR1})-(\ref{PSR2}) agree with
Ref.~\cite{khodja2} in the $m_b\to\infty$ limit.

For the $B\to V$ transitions more work is needed. We have recalculated the
correlator considered in Ref.~\cite{aliev}
\beq
\Pi_\mu(p^\prime,q)=i\int d^4x\,e^{iq\cdot x}
\left\langle V,\,p^\prime\left|\,T\,\overline{q}(x)\gamma_\mu(1-\gamma_5)b(x)\overline{b}(0)i\gamma_5q(0)\,\right|0\right\rangle
\eeq
taking into account the contribution of the distribution amplitudes
$h\dinal^{(s)}(u)$ and $h\dinal^{(t)}(u)$. Using the method and notations of
Ref.~\cite{aliev}, we find~\footnote{Recall our conventions, footnote
\#~\ref{conventions}.}
\beqa
\Pi_\mu(p^\prime,q)=&-&m_bf_Vm_V\int_0^1\frac{du}{\Delta}\,
\left[\epsilon_\mu^\ast g\trans^{(v)}+2(q\cdot\epsilon^\ast)p^\prime_\mu\frac{1}{\Delta}
\left(\Phi\dinal-G\trans^{(v)}\right)\right]\nn\\
&-&i\epsilon_{\mu\nu\rho\sigma}\epsilon^{\ast\nu}p^{\prime\,\rho}q^\sigma
\left[\frac{m_b}{2}f_Vm_V\int_0^1\frac{du}{\Delta^2}\,g\trans^{(a)}
+f_V^\perp\int_0^1 du \frac{\phi\trans}{\Delta}\right]\nn\\
&-&f_V^\perp\int_0^1 du\,\frac{\phi\trans}{\Delta}
\left[\epsilon_\mu^\ast(p^\prime\cdot q+p^{\prime\,2}u)-p_\mu(q\cdot\epsilon^\ast)\right]\nn\\
&-&f_V^\perp m_V^2\int_0^1\frac{du}{\Delta}\,
\epsilon_\mu^\ast\left[\left(1+2\frac{m_b^2}{\Delta}\right)
\left(H\dinal^{(t)}-\Phi\trans\right)+\frac{1}{2}h\dinal^{(s)}\right]\nn\\
&+&2f_V^\perp
m_V^2\int_0^1\frac{du}{\Delta^2}\,(q\cdot\epsilon^\ast)(q_\mu+up^\prime_\mu)
\left(H\dinal^{(t)}-\Phi\trans-\frac{1}{2}h\dinal^{(s)}\right)\,,\label{correl}
\eeqa
the upper-case notation meaning the primitive of a lower-case function:
\beq
F(u)=-\int_0^u dv\,f(v)\,.
\eeq
Compared to Eq.~(19) of Ref.~\cite{aliev}, the last two terms of the above
equation are new~\footnote{They are taken into account in Ref.~\cite{ball4},
together with twist-four and radiative corrections.}.

In taking the Borel transformation of Eq.~(\ref{correl}), a subtlety emerges:
the standard procedure considers the Lorentz-invariant decomposition of
$\Pi_\mu$, that is the coefficients of $\epsilon_\mu^\ast$,
$(q\cdot\epsilon^\ast)p^\prime_\mu$ and $(q\cdot\epsilon^\ast)q_\mu$, and
performs the Borel transformation on the variable $(p^\prime+q)^2$ with
$p^{\prime\,2}=m_V^2$ and $q^2$ fixed to their physical value. This amounts in
particular to consider $q\cdot\epsilon^\ast$ as a constant. While this is true
for transverse mesons, for which $q\cdot\epsilon^\ast\trans=0$, it is clearly
not possible for longitudinal mesons. Thus for longitudinal mesons we simplify
the correlator $\Pi_\mu$ by expressing $q\cdot\epsilon^\ast\dinal$ in terms of
$p^\prime\cdot q$ and $q^2$ before Borel transforming on $(p^\prime+q)^2$:
\beq
q\cdot\epsilon^\ast\dinal=\left(\frac{p^\prime\cdot
q}{m_V}+m_V\right)\sqrt{1-\frac{m_V^2}{E^2}}\,.
\eeq

In the limit $m_b\to\infty$ and $E\to\infty$ the $B\to V$ semileptonic matrix
elements can finally be written under the form~(\ref{B->Vv})-(\ref{B->Va})
with
\beqa
\zeta\dinal^{(a)}(M,E)&=&\frac{1}{f_B}\frac{1}{2E^2}
\left[-f_V\phi\dinal^\prime(1)I_2(\omega_0,\mu_0)+f_V^\perp m_Vh\dinal^{(t)}(1)I_1(\omega_0,\mu_0)\right]\,,\label{VSR1}\\
\zeta_{1/\!/}^{(a)}(M,E)&=&\frac{1}{f_B}\frac{f_V^\perp m_V}{2E^2}
\left[h\dinal^{(t)}(1)+\frac{1}{2}h\dinal^{(s)\,\prime}(1)\right]I_1(\omega_0,\mu_0)\,,\\
\zeta\trans^{(a)}(M,E)&=&\frac{1}{f_B}\frac{1}{2E^2}
\left[-f_V^\perp\phi\trans^\prime(1)I_2(\omega_0,\mu_0)+f_Vm_V g\trans^{(v)}(1)I_1(\omega_0,\mu_0)\right]\,,\label{VSRm}\\
\zeta\trans^{(v)}(M,E)&=&\frac{1}{f_B}\frac{1}{2E^2}
\left[-f_V^\perp\phi\trans^\prime(1)I_2(\omega_0,\mu_0)-\frac{1}{4}f_Vm_V g\trans^{(a)\,\prime}(1)I_1(\omega_0,\mu_0)\right]\,.\label{VSR2}
\eeqa
As for the transverse form factors at the particular point $q^2=0$,
Eqs.~(\ref{VSRm})-(\ref{VSR2}) agree with Ref.~\cite{ball2} in the
$m_b\to\infty$ limit.

Let us now discuss our results, Eqs.~(\ref{PSR1})-(\ref{PSR2})
and~(\ref{VSR1})-(\ref{VSR2}). First, as $f_B\sqrt{m_B}\sim{\rm
C^{st}}$~\cite{IW}, the factorized scaling law $\sim\sqrt{M}/E^2$ is clearly
seen, as anticipated in Section~\ref{zeta}. Note that the $1/E^2$-dependence
holds despite the fact that $\phi_p$, $g\trans^{(v)}$ and $h\dinal^{(t)}$ do
not vanish at $u=0,\,1$, which is a hint that this scaling law may be
independent of the LCSR calculation. Second it still seems that we have six
independent form factors to describe the semileptonic $B\to P(V)$ transitions.
However, using the results of Ref.~\cite{ali,BF,ball3} based on the conformal
expansion of the distribution amplitudes and the equations of motion, the
following relations hold exactly in QCD
\beqa
\phi_p(1)+\frac{1}{6}\phi_\sigma^\prime(1)&=&0\,,\label{fo1}\\
h\dinal^{(t)}(1)+\frac{1}{2}h\dinal^{(s)\,\prime}(1)&=&0\,,\label{fo2}\\
g\trans^{(v)}(1)+\frac{1}{4}g\trans^{(a)\,\prime}(1)&=&0\,.\label{fo3}
\eeqa
Reporting the above relation in Eqs.~(\ref{PSR1})-(\ref{PSR2})
and~(\ref{VSR1})-(\ref{VSR2}) we obtain $\zeta_1=\zeta_{1/\!/}=0$ and that the
semileptonic $B\to P(V)$ matrix elements can finally be written under the
form~(\ref{res1Deb})-(\ref{res1Fin}) with
\beqa
\zeta(M,E)&=&\frac{1}{f_B}\frac{1}{2E^2}
\left[-f_P\phi^\prime(1)I_2(\omega_0,\mu_0)+\frac{f_Pm_P^2}{m_{q_1}+m_{q_2}}\phi_p(1)I_1(\omega_0,\mu_0)\right]\,,\label{resSRdeb}\\
\zeta\dinal(M,E)&=&\frac{1}{f_B}\frac{1}{2E^2}
\left[-f_V\phi\dinal^\prime(1)I_2(\omega_0,\mu_0)+f_V^\perp m_Vh\dinal^{(t)}(1)I_1(\omega_0,\mu_0)\right]\,,\label{resSR2}\\
\zeta\trans(M,E)&=&\frac{1}{f_B}\frac{1}{2E^2}
\left[-f_V^\perp\phi\trans^\prime(1)I_2(\omega_0,\mu_0)+f_Vm_V g\trans^{(v)}(1)I_1(\omega_0,\mu_0)\right]\,.\label{resSRfin}
\eeqa

It is interesting to note that Eqs.~(\ref{fo1})-(\ref{fo3}) can be derived
simply using the LEET projection condition~(\ref{projection}). Indeed the
functions $\phi_p(u)$ and $\phi_\sigma(u)$ are defined by (with
$\phi_\sigma(0)=\phi_\sigma(1)=0$)
\beqa
\frac{f_Pm_P^2}{m_{q_1}+m_{q_2}}\,\phi_p(u)&=&\int\frac{En\cdot dx}{2\pi}e^{-iuEn\cdot x}
\left\langle P\left|\,\overline{q}_1(x)i\gamma_5{\cal A}(x|0)q_2(0)\,\right|0\right\rangle\,,\label{phip}\\
\frac{-f_Pm_P^2}{6(m_{q_1}+m_{q_2})}\,\phi_\sigma^\prime(u)\,n_\mu&=&\int\frac{En\cdot dx}{2\pi}e^{-iuEn\cdot x}
\left\langle P\left|\,\overline{q}_1(x)\sigma_{\mu\nu}\gamma_5n^\nu {\cal A}(x|0)q_2(0)\,\right|0\right\rangle\ \ \ \ \ \ \ \label{phisig}
\eeqa
with $p^\prime_\mu=En_\mu$ the hadron four-momentum, $u$ the momentum fraction
of the quark $q_1$ and ${\cal A}$ the path-ordered gluon operator ensuring the
gauge-invariance of the above matrix elements
\beq
{\cal A}(x|0) = \P\exp\left\{ig\int_0^1 dw\,x_\mu A^\mu(wx)\right\}\,.
\eeq

For $u=1$ one may replace in Eqs.~(\ref{phip})-(\ref{phisig}) $q_1(x)$ by the
effective LEET field $q_{1n}(x)$ with $n\s q_{1n}(x)=0$. Using
$\sigma_{\mu\nu}=i(g_{\mu\nu}-\gamma_\nu\gamma_\mu)$ one immediately gets
Eq.~(\ref{fo1}). Eqs.~(\ref{fo2})-(\ref{fo3}) can be obtained similarly.
Furthermore it can be checked than $1/E$ corrections to LEET generates $1-u$
corrections in the distribution amplitudes and thus vanish for $u=1$.

We would like to make here a last comment, of phenomenological interest. While
we have shown that the LCSR approach satisfies the general relations and
scaling laws among the form factors, the same approach also allows to
calculate some of the deviations to the asymptotic limit $M\to\infty$ and
$E\to\infty$. As a first test, we have checked using the most recent
calculations~\cite{ball4} that the relations between the form factors are
quite robust, i.e. they are well satisfied even in the non-asymptotic regime.
However, the $\sim\sqrt{M}/E^2$ scaling law seems to be affected by large
corrections (at $q^2=0$, this is discussed in Ref.~\cite{ali}), which may
appear surprising. We leave this interesting question for further
investigation.

To conclude {\it the Light-Cone Sum Rule method provides an explicit
realization of the HQET/LEET formalism, that is it satisfies the
predictions~(\ref{res1Deb})-(\ref{res3Fin}) exactly in the limit of heavy mass
for the initial meson and large energy for the final one.} This is a
remarkable non-trivial result, and we would like to insist on the extreme
simplicity of Eqs.~(\ref{resSRdeb})-(\ref{resSRfin}).

In addition we show in Ref.~\cite{BT} that the quark models based on the
Bakamjian-\-Thomas formalism also verify the HQET/LEET
relations~(\ref{res1Deb})-(\ref{res1Fin}) between the form factors, as well as
the $\sqrt{M}\,z(E)$ scaling law of Eq.~(\ref{res2}). This is another hint
that the LEET formalism is well adapted to the description of heavy-to-light
transitions.
\section{Phenomenological Discussion}
\label{pheno}
Here our purpose is to write down the standard form factors $f_+$, $A_1$ etc.
in terms of the three universal functions $\zeta$, $\zeta\dinal$ and
$\zeta\trans$, a convenient way to compare our results with previous
approaches, and to discuss some phenomenological applications. The former form
factors are defined as follows:
\beqa
\left\langle P\left|V^\mu\right|B\right\rangle &=&
f_+(q^2)\left[p^\mu+p^{\prime\,\mu}-\frac{M^2-m_P^2}{q^2}q^\mu\right]
+f_0(q^2)\,\frac{M^2-m_P^2}{q^2}q^\mu\,,\label{std1}\\
\left\langle P\left|T^{\mu\nu}q_\nu\right|B\right\rangle &=&
i\frac{f_T(q^2)}{M+m_P}\left[q^2(p^\mu+p^{\prime\,\mu})-(M^2-m_P^2)q^\mu\right]\,,\\
\left\langle V\left|V^\mu\right|B\right\rangle &=&
i\frac{2V(q^2)}{M+m_V}\epsilon^{\mu\nu\rho\sigma}p^\nu
p^{\prime\,\rho}\epsilon^{\ast\sigma}\,,\\
\left\langle V\left|A^\mu\right|B\right\rangle &=&
2m_VA_0(q^2)\frac{\epsilon^\ast\cdot
q}{q^2}q^\mu+(M+m_V)A_1(q^2)\left[\epsilon^{\ast\mu}-\frac{\epsilon^\ast\cdot
q}{q^2}q^\mu\right]\\ &&-A_2(q^2)\frac{\epsilon^\ast\cdot
q}{M+m_V}\left[p^\mu+p^{\prime\,\mu} -\frac{M^2-m_V^2}{q^2}q^\mu\right]\,,\\
\left\langle V\left|T^{\mu\nu}q_\nu\right|B\right\rangle&=&
-2T_1(q^2)\epsilon^{\mu\nu\rho\sigma}p^\nu
p^{\prime\,\rho}\epsilon^{\ast\sigma}\,,\\
\left\langle V\left|T_5^{\mu\nu}q_\nu\right|B\right\rangle&=&
-iT_2(q^2)\left[(M^2-m_V^2)\epsilon^{\ast\mu}-(\epsilon^\ast\cdot
q)(p^\mu+p^{\prime\,\mu})\right]\\ &&-iT_3(q^2)(\epsilon^\ast\cdot
q)\left[q^\mu-\frac{q^2}{M^2-m_V^2}(p^\mu+p^{\prime\,\mu})\right]\,.\label{std2}
\eeqa

Now we consider the matrix elements as given by their asymptotic
expression~(\ref{res1Deb})-(\ref{res1Fin}) with $(v^\mu,n^\mu,E)$
unambiguously defined in Eqs.~(\ref{v})-(\ref{E}); then we identify in these
equations the coefficients of $n^\nu$, $v^\mu$,
$\epsilon^{\ast\mu}-(\epsilon^\ast\cdot v)n^\mu$ etc. with the corresponding
ones in the standard parametrization~(\ref{std1})-(\ref{std2}), keeping all
the light mass terms although, strictly speaking, they are subdominant in the
final hadron large energy limit. The point is that these kinematical mass
corrections could be numerically very large, and thus should not be thrown
away; for example, $M_K^\ast/M_D=0.48$ and $M_K^\ast/M_B=0.17$. Although it
has a certain degree of arbitrariness and introduces some model-dependence,
this procedure amounts to assume that the matrix elements are well
approximated by their asymptotic value~(\ref{res1Deb})-(\ref{res1Fin}), while
the form factors $f_+$, $A_1$ etc. are not, because of the light mass terms
which appear in their definition. This was already postulated in
Ref.~\cite{aleksan}, and this is in rough agreement with
Ref.~\cite{balltwist4}, where it is found that the main light meson mass
corrections are purely kinematical.

We find, with the notation $E_P$ ($E_V$) for the value of $E$ obtained by
putting $m^\prime=m_P$ ($m_V$) in Eq.~(\ref{q2}),
\beqa
f_+(q^2)&=&\zeta(M,E_P)\,,\label{f+}\\
f_0(q^2)&=&\left(1-\frac{q^2}{M^2-m_P^2}\right)\zeta(M,E_P)\,,\\
f_T(q^2)&=&\left(1+\frac{m_P}{M}\right)\zeta(M,E_P)\,,\\
A_0(q^2)&=&\left(1-\frac{m_V^2}{ME_V}\right)\zeta\dinal(M,E_V)+\frac{m_V}{M}\,\zeta\trans(M,E_V)\,,\\
A_1(q^2)&=&\frac{2E_V}{M+m_V}\,\zeta\trans(M,E_V)\,,\label{A1}\\
A_2(q^2)&=&\left(1+\frac{m_V}{M}\right)\left[\zeta\trans(M,E_V)-\frac{m_V}{E_V}\zeta\dinal(M,E_V)\right]\,,\\
V(q^2)&=&\left(1+\frac{m_V}{M}\right)\zeta\trans(M,E_V)\,,\label{V}\\
T_1(q^2)&=&\zeta\trans(M,E_V)\,,\\
T_2(q^2)&=&\left(1-\frac{q^2}{M^2-m_V^2}\right)\zeta\trans(M,E_V)\,,\\
T_3(q^2)&=&\zeta\trans(M,E_V)-\frac{m_V}{E}\left(1-\frac{m_V^2}{M^2}\right)\zeta\dinal(M,E_V)\,.
\label{T3}
\eeqa

Eqs.~(\ref{f+})-(\ref{T3}) make apparent the fact that the form factors $f_0$,
$A_1$ and $T_2$ have a ``kinematical pole'' $\sim (1-q^2/M^2)$ with respect to
the others, which is a finding that was described in Ref.~\cite{aleksan} as
being essentially a relativistic effect. In addition, the $\sim 1/E^2$
dependence of the $\zeta$ form factors, as discussed in Section~\ref{zeta},
imply a dipole behaviour for $f_+$, $f_T$, $A_0$, $A_2$, $V$, $T_1$ and $T_3$
and a pole one for $f_0$, $A_1$ and $T_2$ with the heavy mass as the pole
mass. This pole/dipole description of the form factors was a phenomenological
{\it Ansatz} made in Ref.~\cite{aleksan}. Also Ref.~\cite{aleksan} needed to
introduce some unknown normalization constants, which we may now interpret as
the constants $C$, $C\dinal$ and $C\trans$ of
Eqs.~(\ref{res3Deb})-(\ref{res3Fin}).

Moreover, it becomes clear from Eqs.~(\ref{f+})-(\ref{T3}) that the HQET/LEET
predictions are close to the relations obtained by Stech~\cite{stech} and
Soares~\cite{soares} who have used a constituent quark model approach. Except
some ambiguities in the subleading terms $\sim m^\prime/M$ or $m^\prime/E$,
our general relations~(\ref{f+})-(\ref{T3}) coincide with Stech's and Soares'
ones, if we impose $\zeta\trans=\zeta\dinal$. Note that we have found no
general reason for $\zeta\trans=\zeta\dinal$, and it seems incompatible with
the LCSR explicit expressions~(\ref{resSR2})-(\ref{resSRfin}), where the ratio
$\zeta\trans/\zeta\dinal$, although constant, depends non-trivially on the sum
rule parameters, the decay constants and the light-cone distribution
amplitudes, i.e. on the dynamics. Similarly, in the explicit and covariant
expressions for the form factors that we have obtained in Ref.~\cite{BT} using
the Bakamjian-Thomas quark model approach, it does not seem possible to have
$\zeta\trans=\zeta\dinal$ without any assumption on the quark-quark potential.
Nevertheless, the similarities between Stech's and Soares' predictions and
ours is quite remarkable, and give strong support to these findings.

An interesting phenomenological discussion is done in Ref.~\cite{soaresPheno},
where some tests of the form factor relations, obtained by Stech and Soares,
are performed or proposed. However, some of these applications are not
possible in our case, for example the study of the longitudinal polarization
of the light daughter meson, because the ratio $\zeta\trans/\zeta\dinal$ is
not known from our formalism~\footnote{With $\zeta\trans=\zeta\dinal$, Soares
finds a good agreement between his prediction and the data for
$\Gamma_L/\Gamma_{\rm tot}$ in $B\to K^\ast\,J/\psi$. We feel that in the LCSR
expressions~(\ref{resSRdeb})-(\ref{resSRfin}), the numerical value of the
ratio $C\trans/C\dinal$ is accidentally close to 1, because the decay
constants and the distribution amplitudes are not very different for the
transverse and the vector meson.}. Thus we consider here only the $V/A_1$
ratio, leaving other possible applications for further investigation. From
Eqs.~(\ref{A1}) and~(\ref{V}), this ratio is given by
\beq\label{VsurA1}
\frac{V(q^2)}{A_1(q^2)}=\frac{(M+m_V)^2}{M^2+m_V^2-q^2}\,.
\eeq
The knowledge of this ratio has important consequences: on the one hand it is
measured at $q^2=0$ in the decays $D\to K^\ast\ell\nu_\ell$ and $D_s\to
\phi\ell\nu_\ell$ ~\cite{PDG}; on the other hand it provides the ratio
$\Gamma_+/\Gamma_-$ of the width to the helicity eigenstates $\lambda=\pm 1$.
The latter is also measured in $D\to K^\ast\ell\nu_\ell$~\cite{PDG}, as well
as in $B\to K^\ast\,J/\psi$~\cite{CLEO} where it can be estimated if the
factorization of the non-leptonic matrix elements is
assumed~\cite{aleksan,kramer}.

For the semileptonic decay the ratio $\Gamma_+/\Gamma_-$ reads, thanks to
Eq.~(\ref{VsurA1})
\beq\label{SL}
\frac{\Gamma_+}{\Gamma_-}=\left[\int_0^{E_{\rm max}}
\left|\sqrt{1-m_V^2/E^2}-1\right|^2\,dE\right] /
\left[\int_0^{E_{\rm max}}
\left|\sqrt{1-m_V^2/E^2}+1\right|^2\,dE\right]\,,
\eeq
with
\beq
E_{\rm max} = \frac{M}{2}\left(1+\frac{m_V^2}{M^2}\right)\,,
\eeq
while for the non-leptonic decay $B\to V_1\,V_2$ in the factorization
assumption, after simplification by Eq.~(\ref{VsurA1}), it is given
by~\cite{kramer}
\beq\label{NL}
\frac{\Gamma_+}{\Gamma_-}=\frac{\left|1-\sqrt{1-1/x^2}+m_{V_1}/(xm_{V_2})\right|^2}
{\left|1+\sqrt{1-1/x^2}+m_{V_1}/(xm_{V_2})\right|^2}
\eeq
with
\beq\label{x}
x=\frac{M^2-m_{V_1}^2-m_{V_2}^2}{2m_{V_1}m_{V_2}}\,.
\eeq
Note that in the strict $M\to\infty$ and $E\to\infty$ limit (that imply
$x\to\infty$), one has $\Gamma_+=0$ in both cases, which is reminiscent of the
fact that an ultra-relativistic quark produced by the $V-A$ current is purely
left-handed~\cite{soaresPheno}; the HQET/LEET relation~(\ref{VsurA1}) implies
that the naive picture at the quark level still holds at the hadron level.

The predictions~(\ref{SL}) and~(\ref{NL}) are compared with experimental data
in Table~\ref{table1}. The agreement is striking; as for the ratio
$\Gamma_+/\Gamma_-$ in $D\to K^\ast$, this might be accidental because the
non-zero value of this ratio is obtained from the large, although formally
subleading, kinematical terms in $m_V$, the reliability of which is not clear,
as stressed above. Moreover the decay $D\to K^\ast$ is naively quite far from
the $M\to\infty$ and $E\to\infty$ limit, and the relation~(\ref{VsurA1}) is
assumed quite arbitrarily to hold in the whole range of $q^2$, otherwise the
integration in Eq.~(\ref{SL}) could not be performed. Nevertheless these
results are encouraging and appeal to investigate further the applications of
the HQET/LEET formalism in heavy-to-light decays.
\begin{table}[ht]\begin{center}
\begin{tabular}{|c|ccc|}\hline
Observable & $m_V=0$ & Eq.~(\ref{SL}) or~(\ref{NL}) & Exp. data \\\hline\hline
\multicolumn{4}{|c|}{$D\to\rho\ell\nu_{\ell}$}\\\hline
$V/A_1$ at $q^2=0$ & 1 & 1.70 & - \\ $\Gamma_+/\Gamma_-$ & 0 & 0.11 & -
\\\hline\hline
\multicolumn{4}{|c|}{$D\to K^\ast\ell\nu_{\ell}$}\\\hline
$V/A_1$ at $q^2=0$ & 1 & 1.78 & 1.85 $\pm$ 0.12~\cite{PDG} \\
$\Gamma_+/\Gamma_-$ & 0 & 0.15 & 0.16 $\pm$ 0.04~\cite{PDG} \\\hline\hline
\multicolumn{4}{|c|}{$D_s\to\phi\ell\nu_{\ell}$}\\\hline
$V/A_1$ at $q^2=0$ & 1 & 1.82 & 1.5 $\pm$ 0.5~\cite{PDG} \\
$\Gamma_+/\Gamma_-$ & 0 & 0.18 & - \\\hline\hline
\multicolumn{4}{|c|}{$B\to\rho\ell\nu_{\ell}$}\\\hline
$V/A_1$ at $q^2=0$ & 1 & 1.29 & - \\ $\Gamma_+/\Gamma_-$ & 0 & 0.02 & -
\\\hline\hline
\multicolumn{4}{|c|}{$B_s\to K^\ast\ell\nu_{\ell}$}\\\hline
$V/A_1$ at $q^2=0$ & 1 & 1.32 & - \\ $\Gamma_+/\Gamma_-$ & 0 & 0.02 & -
\\\hline\hline
\multicolumn{4}{|c|}{$B\to K^\ast\,J/\psi$ (factorization)}\\\hline
$\Gamma_+/\Gamma_-$ & 0 & 0.005 & 0.03 $\pm$ 0.08~\cite{CLEO} \\\hline\hline
\multicolumn{4}{|c|}{$B_s\to \phi\,J/\psi$ (factorization)}\\\hline
$\Gamma_+/\Gamma_-$ & 0 & 0.007 & - \\\hline
\end{tabular}
\caption{\it Predictions of HQET/LEET for the ratio $V/A_1$ and the ratio
$\Gamma_+/\Gamma_-$ of the width to the helicity eigenstates $\lambda=\pm 1$
in various decays. The second column quotes the strict $M\to\infty$ and
$E\to\infty$ limit, obtained by putting $m_V=0$ in
Eqs.~(\ref{VsurA1})-(\ref{x}); the third is the result which incorporates the
$m_V\ne 0$ kinematical mass corrections (Eq.~(\ref{SL}) or~(\ref{NL})), as
explained in the text. As for the non-leptonic decays, naive factorization is
assumed, along the line of Refs.~\cite{aleksan,kramer}.\label{table1}}
\end{center}\end{table}

Finally, let us stress that the general relations that we have found among the
form factors could be very useful for extracting the CKM matrix elements. It
has already been shown that the Isgur-Wise relations between the penguin and
semileptonic form factors may allow the extraction of $|V_{ub}|$~\cite{BD}, by
looking at $B\to K^\ast\gamma$ and $B\to\rho\ell\nu_\ell$. Here we have much
more constraints on the form factors, as only one function describe all the
$B\to V\trans$ transitions. Therefore, a reanalysis of the phenomenological
methods already proposed in the literature to extract the CKM couplings could
be very interesting in this respect.
\section{Conclusion}
We have argued that the HQET/LEET formalism seems to be well adapted to the
description of heavy-to-light transitions in the large recoil region. In the
asymptotic limit of heavy mass $M$ for the initial meson and large energy $E$
for the final one, there are only three independent form factors describing
all the ground state heavy-to-light weak current matrix elements. Moreover, a
factorization formula $\sim\sqrt{M}\,z(E)$ is obtained, and a dipole scaling law
$\sim 1/E^2$ should come from the usual expectation of the $\sim(1-u)$
behaviour for the suppression of the Feynman mechanism. We have checked
explicitly that the Light-Cone Sum Rule method verifies these constraints, and
predicts very simple analytical expressions for the form factors. Finally,
there is a first agreement with available experimental data, although more
observables are needed to make definite conclusions.

It is clear that there is a lot of work to do in this field. From the
theoretical point of view, one should establish the HQET/LEET formalism more
firmly than we have done. In particular, the radiative corrections should be
handled. Some interesting questions also concern the relations between the
LEET Lagrangian and the light-cone quantization.

From the phenomenological point of view, one may look at some observables that
are fully predictable in the final hadron large energy limit. The question of
how to treat the main corrections is open. Finally it is tempting to search
for new methods allowing to extract the CKM matrix elements, with the
possibility of controlling the theoretical uncertainties thanks to the general
constraints on the form factors.


\begin{thebibliography}{99}
\bibitem{PDG} Particle Data Group (C. Caso {\it et al.}), Eur. Phys. J. C. {\bf C3}, (1998).
\bibitem{IW} M. Isgur and M. B. Wise, Phys. Lett. {\bf B232}, 113 (1989); {\bf B237},
(1990) 527; Phys. Rev. {\bf D42}, 2388 (1990).
\bibitem{CZ} V. L. Chernyak and I. R. Zhitnitsky, Nucl. Phys. {\bf B345}, 137 (1990).
\bibitem{BD} G. Burdman and J. F. Donoghue, Phys. Lett. {\bf B270}, 55 (1991).
\bibitem{DG} M. J. Dugan and B. Grinstein, Phys. Lett. {\bf B255}, 583 (1991).
\bibitem{khodja1} V. M. Belyaev, A. Khodjamirian and R. R\"uckl, Z. Phys. {\bf C60}, 349 (1993).
\bibitem{ali} A. Ali, V. M. Braun and H. Simma, Z. Phys. {\bf C63}, 437 (1994).
\bibitem{ball2} P. Ball and V. M. Braun, Phys. Rev. {\bf D55}, 5561 (1997).
\bibitem{aliev} T. M. Aliev, A. \"Ozpineci and M. Savci, Phys. Rev. {\bf D56}, 4260 (1997).
\bibitem{aliev2} T. M. Aliev, H. Koru, A. \"Ozpineci and M. Savci, Phys. Lett. {\bf B400}, 194 (1997).
\bibitem{khodja3} A. Khodjamirian, R. R\"uckl, S. Weinzierl and O. Yakovlev, Phys. Lett. {\bf B410}, 275 (1997).
\bibitem{ball1} E. Bagan, P. Ball and V. M. Braun, Phys. Lett. {\bf B417}, 154 (1998).
\bibitem{ball4} P. Ball, J. High Energy Phys. {\bf 9809}, 005 (1998); P. Ball and V. M. Braun, Phys. Rev. {\bf D58}, 094016 (1998).
\bibitem{khodja2} A. Khodjamirian, R. R\"uckl and C. W. Winhart, Phys. Rev. {\bf D58}, 054013 (1998).
\bibitem{BT} J. Charles {\it et al.}, LPTHE-Orsay 98-78, {\tt hep-ph/9901378}.
\bibitem{BT0} A. Le Yaouanc, L. Oliver, O. P\`ene and J.-C. Raynal, Phys. Lett. {\bf B365}, 319 (1996).
\bibitem{agliettiNew} U. Aglietti and G. Corb\`o, {\tt hep-ph/9712242}; Phys. Lett. {\bf B431}, 166 (1998).
\bibitem{rome} U. Aglietti {\it et al.}, Phys. Lett. {\bf B432}, 411 (1998);
{\bf B441}, 371 (1998).
\bibitem{agliettiOld} U. Aglietti, Phys. Lett. {\bf B292}, 424 (1992).
\bibitem{martinelli} O. P\`ene thanks G. Martinelli for a similar argument.
\bibitem{neubert} M. Neubert, Phys. Rept. {\bf 245}, 259 (1994).
\bibitem{BL} See Appendix A of S. J. Brodsky and G. P. Lepage, Phys. Rev. {\bf D22}, 2157 (1980).
\bibitem{BF} V. M. Braun and I. B. Filyanov, Z. Phys. {\bf C48}, 239 (1990).
\bibitem{su2} The SU(2) global symmetry of LEET was already noted in Ref.~\cite{DG},
although the generators were not given explicitly.
\bibitem{georgi} H. Georgi, Phys. Lett. {\bf B240}, 447 (1990).
\bibitem{ball3} P. Ball, V. M. Braun, Y. Koike and K. Tanaka, Nucl. Phys. {\bf B529},  323 (1998).
\bibitem{balltwist4} P. Ball and V. M. Braun, {\tt hep-ph/9810475}.
\bibitem{CZphysRep} V. L. Chernyak and I. R. Zhitnitsky, Phys. Rept. {\bf 112}, 173 (1984).
\bibitem{stech} B. Stech, Phys. Lett. {\bf B354}, 447 (1995); see also the model of
M. Neubert and B. Stech, {\tt hep-ph/9705292}, in {\it Heavy Flavours II}, p. 294,
eds. A. J. Buras and M. Lindner, World Scientific, Singapore (1997).
\bibitem{soares} J. M. Soares, Phys. Rev. {\bf D54}, 6837 (1996); J. M. Soares, {\tt hep-ph/9810402}.
\bibitem{gregory} G. P. Korchemsky and G. Sterman, Phys. Lett. {\bf B340},
96 (1994); A. G. Grozin and G. P. Korchemsky, Phys. Rev. {\bf D53}, 1378
(1996).
\bibitem{aleksan} R. Aleksan {\it et al.}, Phys. Rev. {\bf D51}, 6235 (1995).
\bibitem{soaresPheno} J. M. Soares, {\tt hep-ph/9810421}.
\bibitem{CLEO} The CLEO Collaboration (C. P. Jessop {\it et al.}), Phys. Rev. Lett. {\bf 79}, 4533 (1997); following Ref.~\cite{soaresPheno}, the CLEO
results that are expressed in the transversity basis are translated into the
helicity basis.
\bibitem{kramer} G. Kramer and W. F. Palmer, Phys. Rev. {\bf D45}, 193 (1992).
\end{thebibliography}
\end{document}